# A Moving Three Level Lambda Rubidium Atom Interacting with a Single Mode Cavity Field in the Presence of Kerr Medium

Ahmed Salah[1] and N. H. Abd El-Wahab[2]
[1]Mathematics and Theoretical Physics Department, Nuclear Research Center
Egyptian Atomic Energy Authority, Cairo, Egypt
[2]Mathematics Department, Faculty of Science, Minia University, Minia, Egypt

## Abstract

We study the interaction between a moving three-level lambda rubidium atom and a single mode cavity field in the presence of Kerr-like medium. We derive the basic equations of motion for atomic system and show that it is exactly solvable in the rotating wave approximation (RWA). The momentum increment, the momentum diffusion and the Mandel Q parameter of this system are studied. We investigate numerically the influence of Kerr-Like medium on the evolution of the previous statistical properties in the exact resonate and nonresonate case where the atom is initially prepared in a momentum eigenstate and the field in the squeezed state. It is found that the Kerr medium has an important effect on their evolution. Finally, conclusion and discussion are given.

## 1. Introduction

Many schemes in quantum optics, which is usually used for the description of the atom-field interaction. The simplest one of such schemes is to investigate the atom–field entanglement in the Jaynes–Cummings model (JCM) [1]. The model as well known describes an interaction of a two level atom with a single mode quantized radiation field. In fact the model is of fundamental importance in the field of quantum optics [2, 3] which is realizable to a good approximation in experiments with Rydberg atoms in high-Q superconducting cavities [4]. Indeed, the JCM and its generalizations can predict new physical results. JCM with intensity-dependent coupling suggested by Buck and Sukumar [5,6] describes the dependence of atom-field coupling on the light intensity. This model has been extended in many directions. One of these directions is in addition of further levels; for example, adding a third level is necessary to investigate the physical phenomena associated with the two-photon process [7]. Two exactly solvable generalizations of the JCM have been proposed by Sukumar and Buck [8], one involving intensity dependent coupling and the other involving multiphoton interaction between the field and atom. Also, possible generalizations are the consideration of multimode and multiphoton instead of single mode and single photon [9], addition of Kerr-like medium, and Stark shift [10] have been performed. Also, extensive studies of a three-level atom with different configurations under RWA interacting with quantized fields inside an ideal cavity were carried out in detail by Yoo and Eberly [11]. Later on several more studies on dynamical evolution and field statistics were reported on the similar type of models [12]. These models could be experimentally tested by utilizing three-level atoms in various configurations in the micromaser systems. Based on the JCM and its various extensions, a multitude of interesting nonclassical effects, such as the collapse and revival of Rabi oscillations, antibunched light, squeezing, inversion less light amplification, electromagnetic induced transparency, etc., have been extensively studied [8,12].

A radiation field with nonclassical characters that cannot be explained by classical wave theory is called nonclassical light. The sub-Poissonian photon statistics is one of the observable nonclassical characters of light. Since Mendel's Q-parameter was introduced to examine the photon statistics in resonance fluorescence [13], it has been provided a useful criterion to distinguish a nonclassical light from classical one. Furthermore, Hong and Mandel [14,15] generalized the squeezing into higher orders and demonstrated the exhibition of higher order squeezing in the physical processes of degenerate parametric down conversion, harmonic generation, and resonance fluorescence. Lee [16] introduced a higher order criterion in the photon statistics by using an inequality of the factorial moment of the photon number operator. Recently, Kim [17] reported a general method how to obtain the higher moment of photon number for the analysis of the second order sub-Poissonian photon statistics.

In the presence paper, we extend the model studied a moving three-level rubidium atom interacting with a single mode cavity field in the presence of Kerr medium. The cavity mode is coupled to Kerr medium as well as to three level atom. Physically this model may be realized as if the cavity contains two different species of atoms, one of which behaves like a three-level atom and the other like an anharmonic oscillator in the one-mode field of frequency $\Omega$. This model has some optical applications, since this type of nonlinearity may be realized by letting the electromagnetic radiation pass through a nonlinear Kerr medium. Also, one can think of an experiment with a Ryberg atom in a nonlinear Kerr-like cavity. We obtain the wave function and the reduced density matrix of atomic system by solving the time-dependent Schrödinger equation when the atom is initially prepared in a momentum eigenstate and the field in the squeezed state. After the reduced density matrix is calculate, we investigate some statistical aspects of atomic operators such as, the momentum increment, momentum diffusion and the Mandel's Q parameter.

The plan of the paper is as follows: in section 2 we describe the model and obtain both the constants of motion and the wave function when the atom is initially prepared in superposition state. In section 3, we calculate the momentum increment, momentum diffusion. The influence of the Kerr-like medium on the evolution of these aspects when the atom is initially prepared in the upper state is investigated numerically. In section 4, we calculate the Mandel' Q parameter of the field. Also, the effect of the Kerr medium on the dynamic of Q parameter is investigated numerically. Finally, conclusions are presented.

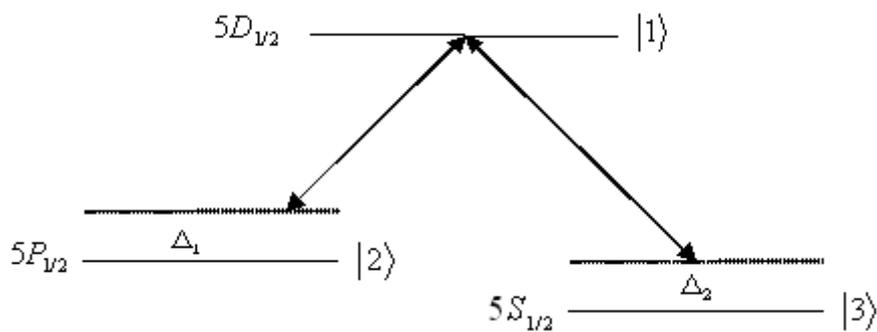

Fig.1. transition schemes of a three-level rubidium atom $|j\rangle$; $j=1,2,3$ interaction with a single mode field with Rabi frequency $\Omega$.

## 2. Model and the Atomic Reduce Density matrix

We consider a moving three-level atom interacting with a single model cavity field in the presence of Kerr-Like medium. As a specific example, we focus on a gas of rubidium (Rb) atoms [18]. The energy levels $5S_{1/2}, 5P_{1/2}$ and $5D_{1/2}$ of Rb can be suitably used to generated the three configuration. The atomic system as describe in Fig.1, consists of three levels denoted by $|j\rangle$ with energies $\omega_j$; $j = 1,2,3$ and a field mode with frequency and the annihilation (creation) boson operator in the field mode $\hat{a}(\hat{a}^\dagger)$. In the RWA, this atomic system can be described by the following Hamiltonian (with $\hbar = c = 1$)

$$\hat{H} = \hat{H}_0 + \hat{H}_I, \tag{1}$$

where the atom-field Hamiltonian $\hat{H}_0$ is defined as:

$$\hat{H}_0 = \frac{\hat{P}^2}{2M} + \sum_{j=1}^{3}\omega_j \sigma_{jj} + \Omega \hat{a}^\dagger \hat{a}, \tag{2}$$

and the interaction Hamiltonian $\hat{H}_I$ is given by :

$$\hat{H}_I = \chi \hat{a}^{\dagger 2}\hat{a}^2 + (\lambda_1 \hat{R}\sigma_{12} + \lambda_2 \hat{R}\sigma_{13} + h.c.), \tag{3}$$

where $\vec{P}$ is the momentum operator and $\hat{R} = \hat{a}e^{i\vec{k}\cdot\vec{r}}$ with $\vec{k}$ and $\vec{r}$ are the propagation vector and the position vector, respectively. The parameter $\chi$ denotes the dispersive part of the third-order nonlinearity of the Kerr medium, $\lambda_\ell$ ($\ell = 1,2$) are the coupling parameter, and $\sigma_{ij} = |i\rangle\langle j|$, $(i = 1,2,3)$ are the atomic transitions operator. The atomic operator ($\sigma_{jk}$) are the generator of the unitary group SU(3) [19]. They commute with the field operator and satisfy the following relation

$$\left[\sigma_{ij}, \sigma_{k\ell}\right] = \sigma_{i\ell}\delta_{jk} - \sigma_{kj}\delta_{\ell i}, \tag{4}$$

where $\delta_{ij}$ is the Kroniker symbol. On the other hand, the field operators satisfy the following relations:

$$\left[\hat{R}, \hat{n}\right] = \hat{R}, \qquad \left[\hat{R}^\dagger, \hat{n}\right] = -\hat{R}^\dagger, \qquad \left[\hat{R}, \hat{R}^\dagger\right] = 1. \tag{5}$$

For the above motion-quantized JC model Hamiltonian (1), we use the Heisenberg equation and obtain, in addition to the following conservations of atomic probability and excitation numbers.

$$\sigma_{11} + \sigma_{22} + \sigma_{33} = \hat{I}, \qquad \hat{a}^\dagger \hat{a} - \sigma_{22} - \sigma_{33} = \hat{N}, \tag{6}$$

a conservation of atomic momentum plus photon momentum of one-mode:

$$\vec{P} + \vec{k}\hat{a}^\dagger \hat{a} = N_0. \tag{7}$$

Using these conservation, the free Hamiltonian for the atomic system can be rewritten as :

$$\hat{H}_0 = \frac{\left(\hat{N} - \vec{k}\hat{N}_0\right)^2}{2M} + \omega_1 \hat{I} + \Omega \hat{N}, \tag{8}$$

and the interaction Hamiltonian now reads

$$\hat{H}_I = \Delta_1 \sigma_{22} + \Delta_2 \sigma_{33} + \chi \hat{a}^{\dagger 2}\hat{a}^2 + (\lambda_1 \hat{R}\sigma_{12} + \lambda_2 \hat{R}\sigma_{13} + h.c.), \tag{9}$$

where the detuning parameters $\Delta_\ell$; $\ell = 1,2$ are given by

$$\Delta_1 = \omega_2 - \omega_1 + \Omega - \frac{\vec{k}\cdot\vec{P}_0}{M} + \frac{k^2}{2M},$$

$$\Delta_2 = \omega_3 - \omega_1 + \Omega - \frac{\vec{k}\cdot\vec{P}_0}{M} + \frac{k^2}{2M}, \tag{10}$$

with the initial momentum $\vec{P}_0$. obviously, these detuning parameters are dependent on the recoil energy $\vec{k}^2/2M$ of the atom and the Doppler shift $\vec{k} \cdot \vec{P}_0/M$. At any time $t > 0$ the wave function $|\psi(t)\rangle$ for the considered atomic system can be written as

$$|\psi(t)\rangle = \sum_{n=0}^{\infty} q_n \left[ A(t) |\vec{P}_0, 1, n\rangle + B(t) |\vec{P}_0 - \vec{k}, 2, n+1\rangle + C(t) |\vec{P}_0 - \vec{k}, 3, n+1\rangle \right]. \quad (11)$$

The coefficients $A(t), B(t)$ and $C(t)$ are the probability amplitudes.

Using the relations $e^{\pm i\vec{k} \cdot \vec{r}} |\vec{P}_0\rangle = |\vec{P}_0 \pm \vec{k}\rangle$, $\hat{a}|n\rangle = \sqrt{n-1}|n-1\rangle$, $\hat{a}^\dagger |n\rangle = \sqrt{n+1}|n+1\rangle$ and $\sigma_{ab}|b\rangle = |a\rangle$, the time-dependent Schrödinger equations $i(d/dt)|\psi(t)\rangle = \hat{H}_I |\psi(t)\rangle$ implies the following system of coupled differential equations

$$i \begin{pmatrix} \dot{A} \\ \dot{B} \\ \dot{C} \end{pmatrix} = \begin{pmatrix} v_1 & g_1 & g_2 \\ g_1 & v_2 & 0 \\ g_2 & 0 & v_3 \end{pmatrix} \begin{pmatrix} A \\ B \\ C \end{pmatrix}, \quad (12)$$

where

$$\begin{aligned} v_1 &= \chi n(n-1), \\ v_2 &= \Delta_1 + \chi n(n+1), \\ v_3 &= \Delta_2 + \chi n(n+1), \\ g_\ell &= \lambda_\ell \sqrt{n+1}, \end{aligned} \quad (13)$$

The solution of the coupled system (12) depends on the initial conditions of the atomic system. We consider that at time $t = 0$, the atom is initially prepared in a squeezed superposition state of the upper and ground state, which can be written as

$$|\psi(t=0)\rangle = \sum_{n=0}^{\infty} q_n \left[ \cos(\theta)|1, n\rangle + \sin(\theta)e^{i\phi}|3, n+1\rangle \right], \quad (14)$$

where $\Psi$ is the atomic phase space of two levels. For $\theta = 0$, the atom is initially prepared into excited state while for $\theta = \pi/2$ the atom is initially prepared in ground state. Also, we consider the field to be in the squeezed state

$$P(n) = \frac{1}{n! \cosh r} \left( \frac{\tanh r}{2} \right)^n \left| H_n(\beta/\sqrt{\sinh 2r}) \right|^2 \exp\left\{ -|\beta|^2 + \frac{1}{2}(\beta^2 + \beta^{*2}) \tanh r \right\}, \quad (15)$$

where $\beta = \alpha \cosh r + \alpha^* \sinh r$ with $\alpha = |\alpha|e^{i\xi}, r \geq 0, \xi$ is the angle between the coherence excitation direction and the squeezing direction.

Using these initial conditions, we can follow the approach of Li et al, [20] and after somewhat lengthy calculations; we find that the solution of the pervious coupled system is given by

$$A(t) = -\sum_{j=1}^{3} C_j (\mu_j + v_2) e^{i\mu_j t},$$

$$B(t) = \sum_{j=1}^{3} \frac{C_j}{g_1 g_2} \left[ \mu_j^2 + (v_2 + v_3)\mu_j + v_2 v_3 - g_2^2 \right] e^{i\mu_j t}, \quad (16)$$

$$C(t) = \sum_{j=1}^{3} C_j e^{i\mu_j t},$$

where $\mu_j$ are the roots of the following third order equation

$$\mu^3 + x_1\mu^2 + x_2\mu + x_3 = 0, \tag{17}$$

where

$$\begin{aligned} x_1 &= v_1 + v_2 + v_2, \\ x_2 &= v_1 v_2 + v_1 v_3 + v_2 v_3 - g_1^2 - g_2^2, \\ x_2 &= v_1 v_2 v_3 - g_1^2 v_3 - g_2^2 v_2, \end{aligned} \tag{18}$$

The general expressions for these roots are given by

$$\mu_j = -\frac{1}{3}x_1 + \frac{2}{3}\sqrt{x_1^2 - 3x_2}\cos\left(\xi + \frac{2}{3}(j-1)\pi\right),\ j = 1,2,3 \tag{19}$$

with

$$\xi = \frac{1}{3}\cos^{-1}\left(\frac{9x_1 x_2 - 2x_1^3 - 27x_3}{2(x_1^2 - 3x_2)^{3/2}}\right),$$

The coefficients $C_j$ are given by

$$C_j = \frac{\Gamma_1(\mu_i + \mu_k) + \Gamma_2 \mu_i \mu_k}{\mu_{ji}\mu_{jk}}, \tag{20}$$

where

$$\begin{aligned} \Gamma_1 &= g_2 \cos(\theta) + v_3 \Gamma_1, \\ \Gamma_2 &= \sin(\theta)\exp(-i\phi), \end{aligned} \tag{21}$$

with $\mu_{jk} = \mu_j - \mu_k$ and $i \neq j \neq k = 1,2,3$.

Having obtained the probability amplitudes $A$, $B$ and $C$ are given by Eq. (16), the wave function for the combined atom-field system is obtained. The reduced density matrix of the field is now easily obtained as

$$\rho_f(t) = |\psi(t)\rangle\langle\psi(t)|, \tag{22}$$

where the state vector is given by Eq. (11) with the probability amplitudes (16). Thus the reduced density matrix of the field can be written as

$$\rho_f(t) = |S\rangle\langle S| + |T\rangle\langle T| + |U\rangle\langle U|, \tag{23}$$

where

$$\begin{aligned} |S\rangle &= \sum_n q_n A |n\rangle, \\ |T\rangle &= \sum_n q_n B |n+1\rangle, \\ |U\rangle &= \sum_n q_n C |n+1\rangle. \end{aligned} \tag{24}$$

With the reduced density matrix calculated, we are therefore in a position discuss some statistical properties of the atom and the field.

## 3. Momentum Increment and Momentum Diffusion

Once the density matrix has been obtained, the mean value of a dynamical operator $\hat{O}$ is defined by $\langle\hat{O}\rangle = Tr\rho_f(t)\hat{O}$. Using this definition and the pervious results, the atomic momentum increment $\langle\Delta\vec{P}\rangle = \langle\vec{P}\rangle - \vec{P}_0$ and the momentum diffusion $\langle(\Delta\vec{P})^2\rangle = \langle\vec{P}^2\rangle - \langle\vec{P}\rangle^2$ can be calculated. We find that

$$\langle \Delta \vec{P} \rangle = -k \left[ \langle T | T \rangle + \langle U | U \rangle \right], \quad (25)$$

and

$$\langle (\Delta \vec{P})^2 \rangle = |k|^2 \left[ \langle T | T \rangle + \langle U | U \rangle - (\langle T | T \rangle + \langle U | U \rangle)^2 \right]. \quad (26)$$

Let us investigate numerically the evolution of the momentum increment and momentum diffusion and see the effect of Kerr medium. We consider the coupling constants $\lambda_1 = \lambda_2 = \lambda$. In our evaluations, we choose the initial conditions of the system as follows: the mean photon number $|\alpha|^2 = 5, \sinh^{-1}(1), \xi = 0$. We shall concentrate on the case $\theta = 0.0$, i.e. when the atom is initially in $|1\rangle$, and take the relative phase $\phi = 0.0$. The numerical results are presented in Fig.2, where we plot the momentum increment and momentum diffusion versus the scaled time $\lambda t$. Also, in this figure, we consider the exact resonant case $\Delta_1 = \Delta_2 = 0.0$. Left plots correspond to the momentum increment and right plots are for the momentum diffusion. The plot in Fig.2 (a) corresponds to the case when the Kerr medium $\chi = 0.001$; Fig.2 (b)'s corresponds to the case when $\chi = 0.2$; Fig.2(c) $\chi = 0.6$. We observe that the evolution of both the momentum increment and the momentum diffusion exhibit collapse and revival phenomena. The influence of the Kerr medium increases the minimum values of momentum increment, whereas it reduces the maximum values of momentum diffusion. Also, as the Kerr medium parameter increases, the collapse-revival period in both evolution decreases, implying more periodically in their evolutions.

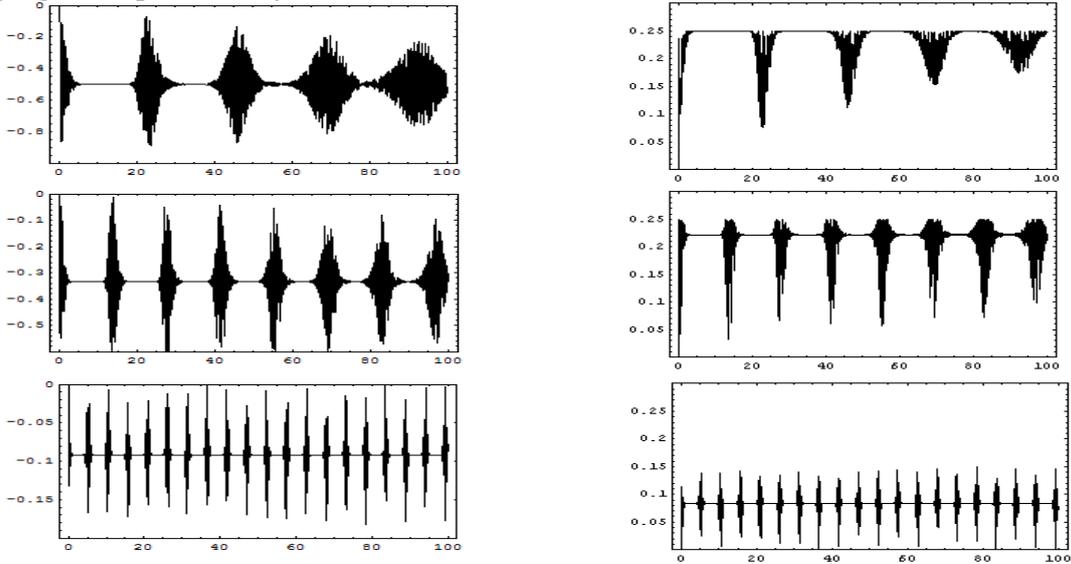

Fig.2. Plot of the momentum increment (left) and the momentum diffusion (right) as a function of scaled time $\lambda t$ with $\Delta_1 = \Delta_2 = 0$,

## 4. Mandel Q Parameter

The Sub-Poissonian photon statistics of the state is one of the most remarkable nonclassical effects. To analyze such effect we study the Mandel Q parameter given by [14, 15]

$$Q(t) = \frac{\langle (\Delta \hat{n})^2 \rangle - \langle \hat{n} \rangle}{\langle \hat{n} \rangle}, \quad (27)$$

where the variance $\langle(\Delta\hat{n})^2\rangle = \langle\hat{n}^2\rangle - \langle\hat{n}\rangle$. For Poissonian statistics, Q = 0. If $Q<0$, the statistics is said to be sub-Poissonian, otherwise, $Q>0$ it is super-Poissonian. Using the wave function of the considered atomic system, we obtain the expectation values $\langle\hat{n}^r\rangle$ and find that

$$\langle\hat{n}^r\rangle = \sum_{n=0}^{\infty} P(n)\left[ n^r |A(t)|^2 + (n+1)^r |B(t)|^2 + (n+1)^r |C(t)|^2 \right]. \quad (28)$$

Inserting Eq.(28) into Eq.(27), the Mandel Q parameter can be obtained.

Let us analyze the effect of the Kerr medium, on the evolution of the Mandel Q parameter as a function of the scaled time $\lambda t$. Where we consider the same as in Fig.2. The numerical results are presented in Fig. 3. We notice that the fluctuations of the Q parameter settle a bit to the negative values, i.e., the statistics of the cavity field settle to the sub-Poissonian as the Kerr medium parameter increases, the sub-Poissonian is increasing. Implying the Kerr medium eliminate the super-Poissonian. It is remarkable from Fig. 3 that the Kerr medium effects make the fluctuations of the Q parameter settle to the negative values, i.e., the statistics settle to be Sub-Poissonian

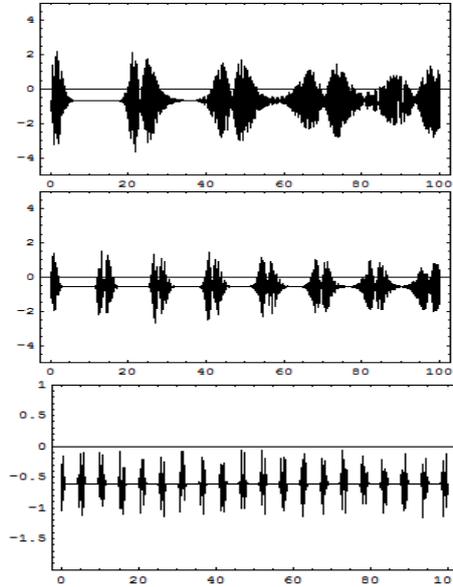

Fig.3. the same as Fig.2 but for Mandel Q parameter

## 5. Conclusions and Discussion

We have studied in this paper an interaction of a three-level rubidium atom with one mode cavity field in the presence of Kerr Medium. We showed that the atomic system is exactly solvable in the RWA and obtained the atom-field wave function when the atom starts in a coherent superposition of the upper and ground states and when the field initially in a squeezed state. The reduced density matrix of the field is also obtained. Various statistical aspects are calculated and the momentum increment, momentum diffusion and the Mandel Q parameter are investigated numerically for the atomic system. The influence of the Kerr-like medium on the evolution of the momentum increment, momentum diffusion and the Mandel Q parameter when the atom is initially prepared in the upper state is analyzed. We observe that the Kerr-like medium have an important effect on the properties of the momentum increment, momentum diffusion and the Mandel Q-parameter.